\documentclass[prl,amsfonts,amssymb,amsmath,floats,twocolumn,aps,footinbib,shownopacs,]{revtex4}

\usepackage[pdftex]{graphicx}
\usepackage{lipsum}
\usepackage{dcolumn}
\usepackage{bm}
\usepackage{color}
\usepackage{physics}
\usepackage[pdftex,colorlinks=true,linkcolor=blue,citecolor=blue,filecolor=blue]{hyperref}

\newcommand{\bi}{\begin{itemize}}
\newcommand{\ei}{\end{itemize}}
\newcommand{\be}{\begin{equation}}
\newcommand{\ee}{\end{equation}}
\newcommand{\ba}{\begin{eqnarray}}
\newcommand{\ea}{\end{eqnarray}}

\newcommand{\imag}{\mathrm{i}}

\begin{document} 

\title{Electronic and fluctuation dynamics following a quench to the superconducting phase}

\author{Christopher Stahl and Martin Eckstein}
\affiliation{Department of Physics, University of Erlangen-Nuremberg, Staudtstra{\ss}e 7, 91058 Erlangen, Germany}
\begin{abstract}
We investigate the dynamics of superconducting fluctuations in the attractive three-dimensional Hubbard model after a quench from the disordered phase to the ordered regime. 
While the long time evolution is well
understood in terms of dissipative time-dependent Ginzburg-Landau models with unstable potentials, early times are 
more demanding
 due to the inseparable dynamics of the pairing fluctuations and the electronic quasiparticles. Our simulation using the time-dependent fluctuation exchange approximation treat both degrees of freedom on the same footing and reveal a non-thermal electronic 
 regime
 causing a non-monotonous growth of the fluctuations. This feature is not directly captured from the Ginzburg-Landau theory, but nevertheless remains observable beyond the thermalization time of the electrons. We further explore how the growth of the order parameter fluctuations leads to an opening of a pseudo-gap in the electronic spectrum, and identify Andreev reflections as the dominant mechanism behind the gap opening.
\end{abstract} 
 
\maketitle 

{\em Introduction --} Ultrafast  pump-probe experiments have demonstrated the possibility to switch between different phases of matter, or even induce new symmetry broken states. To name just a few examples, this includes structural transitions \cite{Wall2018,Storeck2019,Zhou2019},  charge-density wave states \cite{Huber2014, Ishikawa2014, Zong2019, Mitrano2019}, exciton condensates \cite{Mor2017}, and light-induced superconductivity \cite{Fausti2011,Mitrano2016}. State of the art scattering techniques such as diffuse X-ray scattering using free electron lasers, or time-resolved electron diffraction have made it possible to measure time-dependent fluctuations of various order parameters at dynamically induced phase transitions.
Their observation, starting from the very early times, provides the means to tackle two fundamental questions: firstly, how electronic orders emerge out of a disordered state, and second, how the fluctuating short range orders are reflected in the electronic structure.

 A dynamically induced symmetry breaking transition will involve physics on very different timescales. At the earliest times after an electronic excitation, one can expect non-universal dynamics dominated by non-thermal electrons. After the electron thermalization, which is usually assumed to be fast, the dynamics of the order parameter field is described by a time-dependent Ginzburg-Landau theory, with noise and dissipation resulting from electrons in a quasi-thermal state \cite{Hohenberg1977}. This dynamics can give rise to rich phenomena such as pre-thermalization, critical slow down, non-equilibrium scaling behaviors \cite{Nowak2014,Lemonik2017, Lemonik2018, Lemonik2018a,Dolgirev2020}, and even metastability when competing orders are involved \cite{Sun2020}. In the vicinity of the instability, one would generally expect an unstable exponential growth of the order parameter. The final stage of the dynamics is then determined by classical coarsening kinetics \cite{Bray1994}. 

Because electrons thermalize quickly, the early non-thermal regime is often modelled by a quench or fast ramp of the parameters in the effective Ginzburg-Landau theory. However, in the presence of a subsequent unstable growth, it may well be that initial non-universal order parameter fluctuations which are build up at the early stage of the dynamics remain observable even later on. In this paper, we therefore address the crossover in the dynamics from the non-thermal electron  regime to the exponential growth phase at a superconducting transition. This requires  a fully electronic theory which is  beyond (dynamical) mean-field studies of time-dependent symmetry breaking \cite{Chou2017,Sentef2016,Tsuji2013,Bauer2015,Werner2012}, and treats the mutual interaction of electrons with momentum-dependent pairing fluctuations up to times which are sufficiently long compared to the electronic thermalization times.  We find that the pairing fluctuations which are build up in the non-universal initial phase indeed can lead to characteristic  anomalies in the pairing correlations during the exponential growth phase. In particular, this initial phase can give rise to a regime in which order parameter fluctuations show a non-monotonous behavior, with an initial over-population of  modes at momenta above a scale $q_*$. Our analysis also  demonstrates how non-equilibrium pairing correlations become evident early on in the electronic spectrum through Andreev reflection resulting in the opening of a pseudo-gap.

{\em Model and numerical implementation --} We study the three-dimensional attractive one-band Hubbard model,
\begin{equation}
H=\sum_{\bm k\sigma}(\epsilon_{\bm k}-\mu) c^\dagger_{\bm k\sigma}c_{\bm k\sigma}+\frac{U}{N}\sum_{\bm q}\Delta_{\bm q}^\dagger\Delta_{\bm  q}.
\label{eqn:Hubbard}
\end{equation}
Here $c_{\bm k\sigma}^\dagger$ creates an electron with spin $\sigma\in\{\uparrow,\downarrow\}$ in the momentum state $\bm k$. The interaction term is already written in terms of the superconducting order parameter  $\Delta_{\bm q}=\sum_{\bm k}c_{\bm k\uparrow}c_{-\bm k+\bm q\downarrow}$ and $\Delta_{\bm q}^\dagger=\sum_{\bm k} c^\dagger_{-\bm k+\bm q\downarrow}c^\dagger_{\bm k\uparrow}$ for an attractive interaction ($U<0$). For the numerical simulation we assume a continuum limit (electrons in the vicinity of a band minimum), so that the dispersion is $\epsilon_{\bm k}=k^2$, and momentum sums become $(1/N)\sum_{\bm k}=\int d^3k/(2\pi)^3$, with a large momentum cutoff $|\bm  k|<k_c$.  We choose the cutoff $k_c=\pi$ and $\mu=2.59$, so that $k_F=0.57k_c$, and  approximately $18\%$ of the states within the cutoff are filled. We have confirmed that the cutoff is large enough so that resulting errors, such as a violation of the density conservation, are negligible (see appendix).

The non-equilibrium dynamics of the system is solved on the $L$-shaped Keldysh contour, which extends the equilibrium Matsubara formalism to dynamical problems. Using the notation of Ref.~\cite{Aoki2014}, we introduce the contour time-ordered Green's function
\begin{equation}
G_{\bm k\sigma}(t,t')=-\imag\expval{T_{\mathcal{C}}[c_{\bm k\sigma}(t)c_{\bm k\sigma}^\dagger(t')]}
\label{eqn:Green}
\end{equation}
and the propagator for the superconducting fluctuations
\begin{equation}
\chi_{\bm q}(t,t')=-\frac{\imag}{N}\expval{T_\mathcal {C}[\Delta_{\bm q}(t)\Delta^\dagger_{\bm q}(t')]}.
\end{equation}
The latter allows to extract the superconducting fluctuations $C_{\bm q}(t)=\frac{1}{N}\expval{\Delta^\dagger_{\bm q}(t)\Delta_{\bm q}(t)}=-\,\text{Im} \chi_{\bm q}^<(t,t) $. 

We employ a self-consistent fluctuation-exchange approximation (FLEX) \cite{Bickers1989} which expands $\chi_{\bm q}$ in the particle-particle ladder diagrams of the electronic Green's function, corresponding to the dominant divergent channel in equilibrium, and 
takes
into account a self-consistent interaction of  electrons and these  pairing fluctuations. 
By construction, this approach can capture only the normal phase and the exponential growth regime, not the subsequent dynamics well in the symmetry broken phase. The derivation, e.g., using a Hubbard-Stratonovich decoupling of the interaction in the pairing channel,  has been presented in the literature \cite{Engelbrecht2000,Deisz1998,Lemonik2018a}.  For the purpose of the present work, we push the time-dependent FLEX implementation of in Ref.~\cite{Dasari2018} to long times and adapt it for three-dimensional systems. The equations to be solved are 
RPA equations
for $\chi_{\bm q}(t,t')$
\begin{equation}
\chi_{\bm q}(t,t')=\chi^0_{\bm q}(t,t')+\int_\mathcal{C}\!\!\dd \bar{t}\,\chi^0_{\bm q}(t,\bar{t})U(\bar{t})\chi_{\bm q}(\bar{t},t'),
\label{eqn:RPA}
\end{equation}
in terms of the bare propagator
\begin{equation}
\chi_{\bm q}^0(t,t')=
\frac{\imag}{N}\sum_k 
G_{\bm k}(t,t')G_{\bm q-\bm k}(t,t'),
\label{eqn:chi_0}
\end{equation}
and the Dyson equation for the Green's function, $G_{\bm k}(t,t')=[\imag\partial_t-(\epsilon_{\bm k}-\mu)-\Sigma_H(t)-\Sigma_{\bm k}(t,t') ]^{-1}$, with the Hartree self-energy $\Sigma_H(t)=U\expval{n(t)}$; the FLEX self-energy $\Sigma_k$ describes the interaction of electrons with the fluctuating interaction $V_{\bm q}(t,t')=U(t)\chi_{\bm q}(t,t')U(t')$
\begin{equation}
\Sigma_{\bm k}(t,t')=-
\frac{\imag}{N}\sum_{\bm q}
V_{\bm q}(t,t')G_{\bm q-\bm k}(t',t).
\label{eqn:Sigma}
\end{equation}
 This set of equations is solved self-consistently, using the NESSi simulation package \cite{Schueler2020}, with a paralellization over $\bm k$. For the spherical symmetric system, the functions  $G_{\bm k}, \Sigma_{\bm k}$ and $\chi_{\bm q}$ depend only on the absolute value of $k=|\bm k|$, which reduces the required computer memory and makes the present simulations feasible on 400 $\bm k$-points. Momentum integrals 
in \eqref{eqn:chi_0} and \eqref{eqn:Sigma}
can then be rewritten in spherical coordinates. Further details of the numerical implementation are presented in the appendix.
 
\begin{figure}
\includegraphics[width=\linewidth]{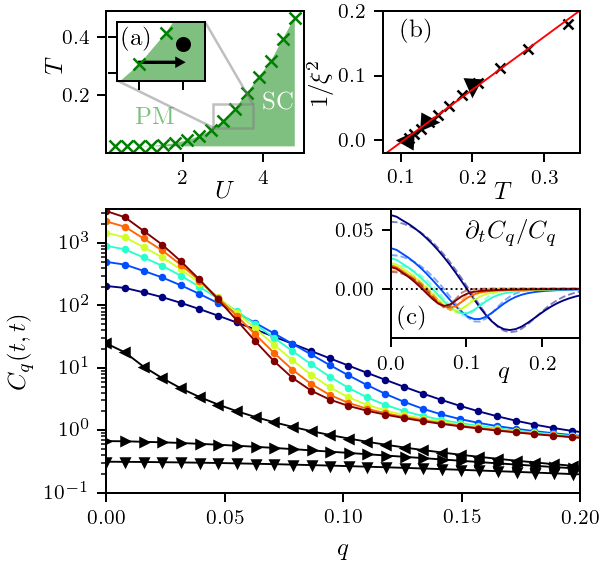}
\caption{(a) Equilibrium phase diagram for the Hubbard model Eq.~\eqref{eqn:Hubbard}. The  arrow indicates the quench, and the black point the assumed final state of the electronic system after thermalization. (b)~Correlation length $\xi(T)$ for $U=3.0$, with a linear fit extracting the critical temperature $T_c$. (c)~Black triangles: Equilibrium correlation $C_q$ for the temperatures indicated with corresponding symbols in (b). The lowest temperature is the initial state for the subsequent dynamics. Solid lines with colored dots: Time-dependent correlations $C_q(t)$ for the quenched system at $t=20,40,\ldots,120$ (blue to red). \textit{Inset:} Normalized derivative $\partial_tC_q(t)/C_q(t)$ (solid) with fits based on Eq.~\eqref{eqn:Correlation} (dashed).}
\label{fig:Chi_time}
\end{figure}

{\em Results: Order parameter growth --}
The black lines in Fig.~\ref{fig:Chi_time}(c) exemplarily show the correlation function $C_q$ in equilibrium for three temperatures. Using a Lorentzian fit $A(\xi^{-2}+q^2)^{-1}+const.$  we determine the correlation length $\xi(T)$.  The correlation length scales like ${|T-T_c|}^{-1/2}$ as expected for the FLEX approximation, 
which shows mean field scaling in equilibrium, see Fig.~\ref{fig:Chi_time}(b). This analysis is used to map out the phase transition line in Fig.~\ref{fig:Chi_time}(a). 
In the symmetry broken phase for $T<T_c$, the FLEX approach does not yield a convergent equilibrium solution. 

To explore the dynamical role of the  
pairing fluctuations
after quenching the system in the unstable regime, the system is prepared in the normal state close to the equilibrium phase transition at $T=0.11>T_c$ and $U=-3.0$, and quenched across the equilibrium phase transition to $U=-3.5$, as indicated by the arrow in Fig.~\ref{fig:Chi_time}(a).  After the quench, a significant increase of the pairing fluctuations can be seen in Fig.~\ref{fig:Chi_time}(c) 
(\textit{blue to red lines}). 
 The correlations are clearly peaked around $q=0$,  indicating the approach of a homogeneous superconducting phase. On the other hand, in the  momentum range $|q|\gtrsim0.05$ the dynamics is clearly  non-monotonous. 
Fluctuations quickly increase at early time, but at later times decrease towards a steady function. 
This is made more clear through the zero-crossing of the normalized derivative $\partial_t C_q(t)/C_q(t)$ at some scale $q_*$ 
(Fig.~\ref{fig:Chi_time}(c), inset).

To analyze this growth of fluctuations, we contrast it with the prediction from a phenomenological classical theory for a complex order parameter field $\phi_q$. A suitable model is {\em model A} according to Hohenberg and Halpherin~\cite{Hohenberg1977}, which describes the dynamics generated by  a Landau-Ginzburg-Wilson Hamiltonian $H_\phi$ for an order parameter field $\phi_{\bm q}(t)$ without coupling to conserved quantities. In the vicinity of the instability we can restrict ourselves to the Gaussian approximation 
$H_\phi(t)=\sum_{\bm q} (r+\ell^2q^2)|\phi_{\bm q}(t)|^2$, 
where $r=0$ sets the phase transition~\cite{Taeuber2014}. The equations of motion are then given by
\begin{align}
\partial_t \phi_{\bm q}(t)
=&-\frac{D}{2}[r(t)+\ell^2q^2]\phi_{\bm q}(t)+\zeta_{\bm q}(t).
\label{eqm_momentum}
\end{align}
Here the
diffusion constant $D$ and the length $\ell$ set the time and length scales, and $\zeta_{\bm q}(t)$ is an Einstein-correlated  white noise characterized by $\expval{\zeta_{\bm q}(t)}=0$ and 
$\expval{\zeta_{\bm q}(t)\zeta_{\bm q'}^*(t')}=DT\delta(t-t')\delta_{{\bm q},{\bm q'}}$, 
which includes the coupling of the order parameter to fast electronic degrees of freedom. We then assume a sudden quench of the $r$ parameter and the bath temperature $T$ to some final values $r_f$ and $T_f$.  Starting from an initial state with correlations $C_q(t_0)$ at a given time $t_0$, the solution gives (see appendix) 
\begin{align}
\label{eqn:Correlation}
C_q(t)=C_q(t_0)e^{-Da_q(t-t_0)}
+\frac{T_f}{a_q}\big[1-e^{-Da_q(t-t_0)}\big ],
\end{align}
where  $a_q=(r_f+\ell^2q^2)$ is used as abbreviation. The first term describes the growth of the initial correlations, while the second part is the noise-driven dynamics. For $r_f<0$, there is an instability, leading to unbound growth of fluctuations at  $q<\sqrt{|r_f|}/\ell$ ($a_q<0$), while the fluctuations relax to a steady form $\frac{T_f}{a_q}$ for larger $q$.

This analysis shows that a quench within  {\em model A}  cannot easily describe the observed behavior in the FLEX simulation. In particular, a zero in the  derivative $\partial_t C_q(t)$ would occur at a scale  $q=q_*$ set by $C_{q_*}(0)=T_f/(r_f+\ell^2 q_*^{2})$; to fulfill this condition with initial equilibrium correlations $C_{q_*}(0)=T_i/(r_i+\ell^2 q_*^{2})$ and  $r_i>r_f$, one would need to make the counter-intuitive approximation that the temperature $T_f$ decreases in the quench $(T_f<T_i)$, and even with that, the scale $q_*$ predicted by  {\em model A} would be time-independent, in contrast to the observation in Fig.~\ref{fig:Chi_time}(c). Consequently, a fit of the FLEX data with the result of Eq.~\eqref{eqn:Correlation} (see dashed lines in the inset of Fig.~\ref{fig:Chi_time}(c)) requires different parameters at each time, and also a strong variation of the parameter $\ell$ and $D$ which are usually kept fixed in the effective model in the vicinity of the phase transition (see the appendix for the fitting parameters).

The explanation is that the effective {\em model A} dynamics cannot be expected to hold for early times, in which electrons are not thermalized, so that time $t_0$ for the onset of the {\em model A} regime is to be set greater than $0$. The  scale $q_*$,  which remains during the gaussian growth phase, is the remanence of the non-thermal correlations which have been build up during the initial phase of the dynamics. In agreement with this, the behavior of $\partial_t C_q(t)/C_q(t)$ at later times ($t\gtrsim60$) becomes increasingly well described by {\em model A}: For small $q$, the normalized derivative $\partial_t C_q(t)/C_q(t)$ approaches a function of the form $-D(r_f+\ell^2 q^{2})$, and the location of the zero-crossing $q=q_*$ becomes time-independent. Moreover, by analysing the 
fluctuation-dissipation  relation of 
the electronic spectra (appendix) we have confirmed that the  electronic degrees of freedom have reached a thermal state with a temperature $T\approx 0.14$ by then. This temperature lies well within the superconducting phase, see black dot in Fig.~\ref{fig:Chi_time}(a).
\begin{figure}
\includegraphics[scale=1]{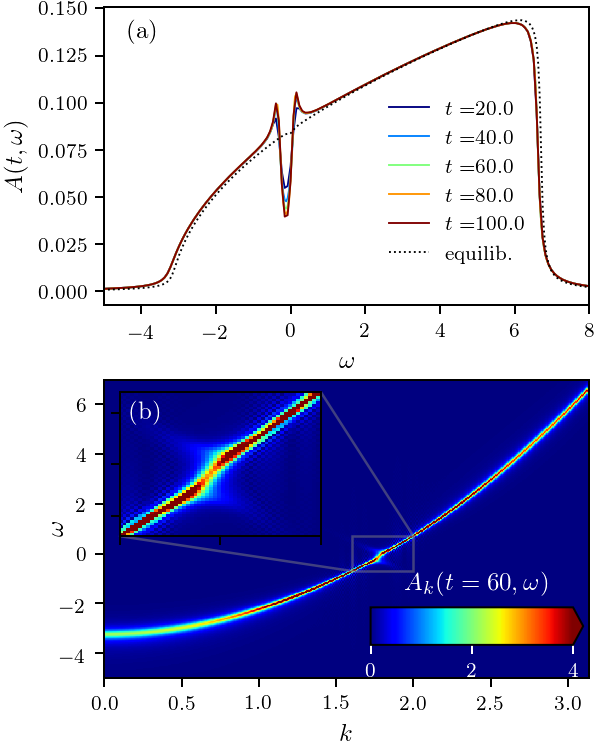}
\caption{(a) Local density of states $A(t,\omega)$ at $t=20,\,40,\,60,\,80,\,100$ inverse hoppings after the quench. The initial equilibrium spectrum is shown by the dotted line. Note that because the density of states depends on energy, the redistribution of weight implies a shift $\Delta\mu=-0.1$ of the chemical potential and thus the location of the pseudo gap. (b) Momentum-dependent spectral function $A_k(t,\omega)$ at $t=60$.}
\label{fig:electronic_obs}
\end{figure}

{\em Electronic spectra --}
In the second part of this analysis, we investigate how the buildup of pairing correlations is reflected in the electronic spectra. The time-dependent spectral weight, obtained from the Wigner-transformation 
\begin{equation}
\label{gehjekle}
A(t,\omega)=-\frac{1}{\pi}\Im\int\!\!\dd s\, G^{ret}(t+s/2,t-s/2)e^{\imag\omega s}
\end{equation}
of the local Green's function 
$G(t,t')=(1/N)\sum_{k} G_k(t,t')$, 
shown in Fig.~\ref{fig:electronic_obs}(a), exhibits the redistribution of spectral weight at the Fermi-edge, opening a pseudo-gap with increasing depth at $\omega=-0.1$. 
Note that spectra are not shown at the maximum simulation time $t=120$, because the vanishing relative time ($s$) range in the Wigner integral \eqref{gehjekle} would limit the frequency resolution; the maximal resolution is obtained at $t=60$.
As the spectral weight $\int\!\dd \omega\, A(t,\omega)=1$ is conserved  at all times, the gap opening is accompanied by the rise of  distinct peaks  above and  below the gap. The momentum dependent spectral function $A_k(t,\omega)$, given by the Wigner-transform of $G^{ret}_k(t,t')$, shows how the dispersion is  depleted at the Fermi-energy, corresponding to the gap opening (Fig.~\ref{fig:electronic_obs}(b)). The peaks next to the pseudo gap in the $k$-integrated spectrum $A(t,\omega)$ are seen to arise from shadow bands above and  below the Fermi-energy (Fig.~\ref{fig:electronic_obs}(b), inset). 

\begin{figure}
\includegraphics[scale=1]{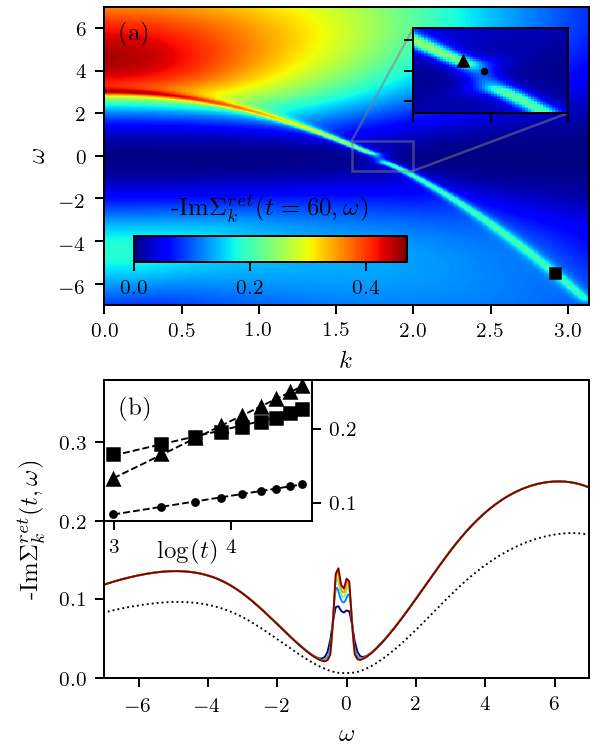}
\caption{(a) Momentum and frequency resolved imaginary part of the self-energy $\Sigma^{ret}_k(t,\omega)$ at $t=60$. (b) Imaginary part of the self-energy at $t=20,40,60,80,100$ (solid, dark-blue, light-blue, green, yellow, red) and the pre-quench equilibrium state (dotted) at the Fermi-momentum. \textit{Inset:} Time dependence of 
$-\text{Im}\Sigma^{ret}_k(t,\omega)$ for different frequency and momenta on the hole dispersion, indicated in (a)
(triangle, dot, square)}
\label{fig:self_energy}
\end{figure}

The underlying mechanism for the gap opening is understood as a consequence of Andreev scattering of electrons on the superconducting fluctuations. Upon scattering with the fluctuations, an electron creates a Cooper pair resonance.
Due to charge and momentum conservation the resonance requires the creation of a hole with the same momentum, that propagates along the trajectory of the electron reversed in time. By inspecting the imaginary part of the self-energy $\Sigma^{ret}_k(t,\omega)$ (which determines the quasi-particle decay rate), we therefore find that a strong resonance around the hole dispersion 
$\omega=-(\epsilon_k-\mu)$ appears in the fluctuating state at long times (Fig.~\ref{fig:self_energy}(a)). 
For electrons at the Fermi-energy, this peak becomes resonant with the quasiparticle energy, and therefore strongly reduces the lifetime: For a momentum on the Fermi-surface, $\text{Im}\Sigma^{ret}_k(t,\omega)$ develops from a Fermi-liquid behavior $\propto \omega^2+const.$  in equilibrium (dashed line) to a peak at $\omega=0$ in the fluctuating phase.  As shown in the inset of 
Fig.~\ref{fig:self_energy}(b) 
the decay rate $\gamma$ of the quasi-particles is increasing proportional to $\log(t)$ at $\omega=0$~(dots), which is well in agreement with analytical findings of Lemonik and Mitra~\cite{Lemonik2017}. 
We find a similar $\log(t)$ behavior for $-\text{Im}\Sigma^{ret}_k(t,\omega)$ for $k$ and $\omega$ on the hole band even away from the Fermi surface, although this of course does not correspond to a quasiparticle lifetime. 

{\em Conclusion --} We derived and implemented a non-local approximation scheme for the three dimensional Hubbard model, to study the growth of pairing fluctuations in a dynamical transition towards a symmetry broken phase. There  are two main results:  (i) 
Although
electrons thermalize quickly, the early non-thermal electron regime can nevertheless have effects 
which remain observable 
at longer times. 
In the present case, the relevant signature is the {\em decrease} of pairing fluctuations in a certain regime where they have been initially overpopulated, while fluctuations at $q=0$ increase.
Related
physics is expected to play a role in charge- and spin-density-wave transitions or exciton condensates, which are described in a similar mathematical way. Clearly, these initial state effects  are not universal and depend on the excitation protocol, but they could be important for the understanding of future time-resolved scattering 
data. It will 
also be interesting to see their role in the dynamics of intertwined orders \cite{Sun2020}, which may depend sensitively on the initial state. 
(ii) Moreover, our simulations show the formation of a pseudo-gap, which is a consequence of the anomalous enhanced decay rate at the Fermi-energy, due to Andreev reflection from pairing fluctuations. 
This demonstrates how even fluctuating (short range) superconductivity can be experimentally observable. A direct scattering measurement of pairing correlations, analogous to charge or spin density wave order parameters, is more challenging, but noise correlation measurements may be an interesting direction \cite{Stahl2019} in this regards.

In future, it would be interesting to go beyond the unstable gaussian regime, and include the higher order diagrammatic corrections which stabilize the order parameter.  
This would finally allow to explore also the subsequent stages of the symmetry breaking dynamics in an electronic model.\\

\begin{acknowledgements}
We thank N. Dasari for discussion and initial collaboration on the implementation, and A. Mitra for useful discussion.
We acknowledge the financial support from the DFG Project 310335100, and the ERC starting grant No.~716648. The numerical calculations have been performed at the RRZE of the University Erlangen-Nuremberg. 
\end{acknowledgements}

\section{Appendix}
\subsection{Details of the numerical implementation} 

The numerical evaluations of the bare propagator $\chi^0_q(t,t')$ and the FLEX self-energy $\Sigma_k(t,t')$ in the three dimensional Hubbard model requires a transformation from 
continuous
Cartesian to spherical coordinates in three dimensions for integrals of the type:
\begin{equation}
A({\bm k})=\int\!\!\dd^3 q\,\, B({\bm q})C({{\bm k} - {\bm q}})
\label{eqn:Cartesian_Integral}
\end{equation} 
with spherical symmetric functions $A$, $B$ and $C$ of the three dimensional momentum vectors $\bm k$ and $\bm q$.
(such as the self energy $\Sigma_{\bm k}$, the Green's function $G_{\bm k}$ or the pairing fluctuations $\chi_{\bm q}$). 
The large momentum cutoff is implicit in these integral, by setting the functions $A({\bm q}) $ and $B({\bm q})$ to zero for arguments $|{\bm q}|>k_c$.
Employing a transformation to spherical coordinates,
with polar angle $\theta$ between $\bm k$ and $\bm q$
yields :
\begin{equation}
A(k)
=\int_0^\infty\!\!\dd q q^2 B(q)\int_{0}^{\pi}\!\!\dd\theta \sin(\theta)\int_0^{2\pi}\!\!\dd\phi C(\abs{{\bm k}-{\bm q}}),
\end{equation}
where $A$ depends only on the absolute value $k=|\bm k|$. 
The argument $\abs{{\bm k}-{\bm q}}$ of the function $C$ is then expressed in polar coordinates as
\begin{equation}
\alpha:=\abs{{\bm k}-{\bm q}}
=
\sqrt{k^2+q^2-2kq\cos{\theta}}.
\end{equation}

\begin{figure}
\includegraphics[scale=1]{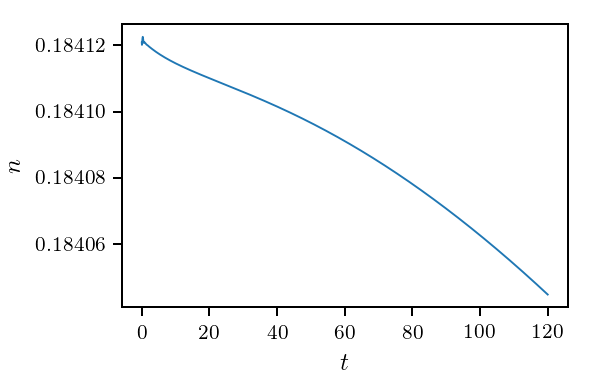}
\caption{Time dependent filling $n(t)$ of the electronic system}
\label{fig:Filling}
\end{figure}

\noindent
We then perform an  integral transformation  from  $\theta$ to $\alpha$, using
\begin{equation}
\dv{\alpha}{\theta}=\frac{kq\sin(\theta)}{\alpha},
\end{equation}
and thus
obtain the final expression for $k\neq 0$:
\begin{align}
A(k)=\frac{2\pi}{k}\int_0^\infty\!\!{\dd} q\,q B(q)\int_{\abs{k-q}}^{\abs{k+q}}\!\!{\dd} \alpha C(\alpha) \alpha.
\end{align}
The limit of this expression at $k=0$ can be obtained from Eq.~\eqref{eqn:Cartesian_Integral} 
by just one spherical transformation as
$A(0)=4\pi\int_0^\infty\!\! \dd q\, q^2B(q)C(q)$.
After application of the transformation and introduction of a momentum cutoff $k_c$ the formula for the self-energy $\Sigma_k$ 
[Eq.~(6) in the main text]
and the bare propagator of the fluctuations $\chi^0_q$ 
[Eq.~(5) in the main text]
read:
\begin{align}
\label{eqn:Sigma_int}
\Sigma_k(t,t')=-\imag\frac{2\pi}{k}\int_0^{k_c}\!\!\dd q\,q V_q(t,t')\int_{\abs{k-q}}^{\abs{k+q}}\!\!\dd \alpha G_\alpha(t',t)\alpha,\\
\label{eqn:Chi_int}
\chi^0_q(t,t')=\imag\frac{2\pi}{k}\int_0^{k_c}\!\!\dd k\,k G_k(t,t')\int_{\abs{k-q}}^{\abs{k+q}}\!\!\dd \alpha G_\alpha(t,t') \alpha,
\end{align}
with $V_q(t,t')=U(t)\chi_q(t,t')U(t')$. 

All the integrals are  calculated using a fifth-order accurate quadrature.
As a technical note, we remark that this requires to handle the inner integral over $\alpha$ from $\abs{k-q}$ to $\abs{k+q}$ with care:
If functions are saved on an equidistant $|k|$ grid, the range of the inner integral extends only over less that five grid points  for small values of $q$ or values close to the cutoff. In order to nevertheless have a fifth-order accurate approximation to the integral, we use a polynomial approximation of the integrand based on grid points outside the integration range.

Further 
note that 
the FLEX approximation is in general one-particle conserving, because it can be derived from a 
Luttinger-Ward-functional. 
However, as a consequence of finite momentum cutoff $k_c$, 
particle number conservation may be violated.
Monitoring the conservation of the one-particle density
\begin{equation}
n(t)=(\frac{4\pi}{3}k_c^3)^{-1}\int_{\abs{\bm k}<k_c}\!\!\!\dd^3 k\,\, G^<_{\bm k}(t,t)
\end{equation}
 for a given cutoff therefore can serve as a heuristic measure that $k_c$ has been chosen sufficiently large.
The numerical 
error due 
to the cutoff $k_c=\pi$ in the main text leads to a loss of less than $0.05\%$ of the initial particles at the end of the calculation, see figure~\ref{fig:Filling}. 

\subsection{Derivation of the model A dynamics} 
\begin{figure}
\includegraphics[scale=1]{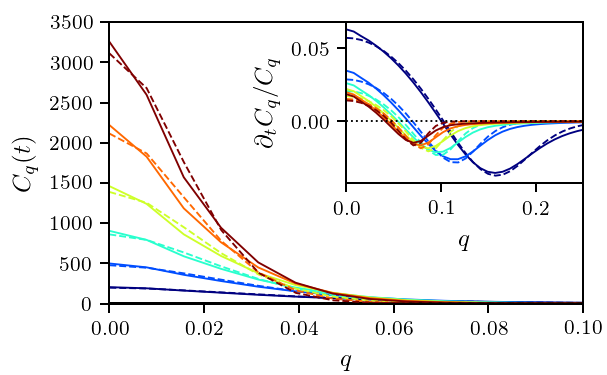}
\caption{Correlation $C_q(t)$ for $t=20,40,\ldots,120$ (solid, blue to red) and the fits based on Eq.\eqref{eqn:Chi}(dashed,blue to red). \textit{Inset:} Normalized derivative of the correlations for the same $t$ (\textit{solid}) and the fits based on the ratio of Eq. \eqref{eqn:Chi_derive} and Eq.\eqref{eqn:Chi}}.
\label{Fitting_model_A}

\includegraphics[scale=1]{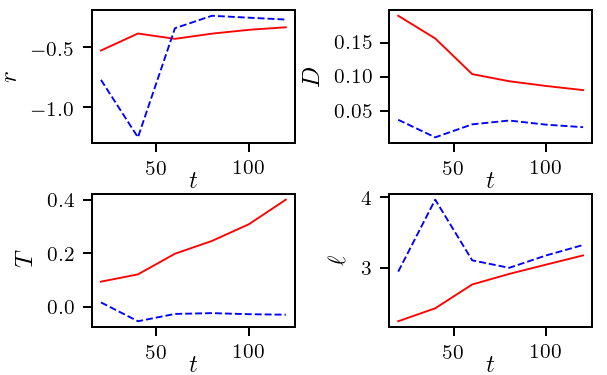}
\caption{Fitting parameters $r$, $D$, $T$, $\ell$ for different time steps for the correlation $C_q$ (solid) and the normalized derivative $\partial_t C_q/C_q$ (dashed)}
\label{Fitting_params}
\end{figure}
The dynamics of model A has been extensively discussed in the literature, for problems in the direct 
vicinity of a critical point, see
Hohenberg and Halperin \cite{Hohenberg1977,Taeuber2014} for a review. We derive here the equation for the 
post-quench 
correlations that was used in the main text.
Model A describes in general the dynamics of a Landau-Ginzburg-Wilson Hamiltonian for the continuous variable of the order parameter field 
$\phi_{\bm q}(t)$ 
without coupling to conserved quantities. In Gaussian approximation, neglecting field-field interaction, the Hamiltonian can be written as $H(t)=
\int d^3\bm q
\big(r(t)+\ell^2q^2\big)|\phi_{\bm q}(t)|^2
$, and 
the dynamics of the order parameter field is given by:
\begin{align}
\begin{split}
\partial_t \phi_{\bm q}(t)=&-\frac{D}{2}\frac{\delta H[\phi_{\bm q}(t)]}{\delta\phi_{\bm q}^*(t)}
+\zeta_{\bm q}(t)\\
=&-\frac{D}{2}[r(t)+\ell^2q^2]\phi_{\bm q}(t)+\zeta_{\bm q}(t)
\label{eqm_momentum}
\end{split}
\end{align}
with the diffusion constant $D$. The Einstein-correlated white noise 
$\zeta_{\bm q}(t)$, characterized by
$\expval{\zeta_{\bm q}(t)}=0$ and $\expval{\zeta_{\bm q}(t)\zeta_{\bm q'}^*(t')}=DT\delta(t-t')\delta_{\bm q,\bm q'}$,
models the interaction of the superconducting fluctuations $\phi_{\bm q}$ with the electronic degrees of freedom.
The general solution of the differential equation is given by
\begin{align}
\begin{split}
\phi_{\bm q}(t)=&\phi_{\bm q}(t_0)e^{-\frac{D}{2}\int_{t_0}^t\dd t'(r(t')+q^2)}\\
&+\int_{t_0}^t\dd t''\zeta_{\bm q}(t'')e^{-\frac{D}{2}\int_{t''}^{t}\dd t'(r(t')+q^2)}.
\label{eqn:solution}
\end{split}
\end{align}
This expression is then inserted in the expectation value $C_{q}(t)=\expval{\phi_{\bm q}(t)\phi_{\bm q}^*(t)}$ in order to calculate the correlations $C_{q}$ (which depend only on $q=|\bm q|$). Assuming a given state with correlations $C_{q}(t_0)$ at a time $t=t_0$,  after which the system is evolved with a given potential $r_f$ and a temperature quench to $T_f$, one obtains for $t>t_0$,
\begin{align}
\begin{split}
C_{q}(t)=&C_{q}(t_0)e^{-D(r_f+\ell^2q^2)(t-t_0)}\\ 
&+\frac{T_f}{(r_f+\ell^2q^2)}[1-e^{-D(r_f+\ell^2q^2)(t-t_0)}].
\label{eqn:Chi}
\end{split}
\end{align} 
This equations can be read in two ways:

(i) For a stable potential $r>0$, the initial correlations eventually damp out, showing that the correlations in equilibrium are of the form $C_{q}=\frac{T}{(r+\ell^2q^2)}$, which is proportional to $\frac{1}{(\xi^{-2}+q^2)}$ with the correlation length $\xi=\ell/\sqrt{r}$. This is the form (with a background) used to fit the equilibrium FLEX data in the main text.

(ii) Equation~\eqref{eqn:solution} for $t>t_0$ can be used after a quench to the unstable potential with $r_f<0$ and a simultaneous temperature quench to $T_f$. This corresponds to Eq.~(8) of the main manuscript. For example, such a  quench could be from a  stable potential $r_i>0$ to $r_f<0$, with $r(t)=r_i+(r_f-r_i)\theta(t)$, and $T(t)=T_i+\theta(t)(T_f-T_i)$, but also any other initial time $t_0$ with initial correlations $C_q(t_0)$ can be used in Eq.~\eqref{eqn:Chi}.

The temporal derivative of the correlations is given by:
\begin{align}
\partial_t C_{q}(t)
=&D[T-(r_f+\ell^2q^2)C_{q}(t_0)]e^{-
D(r_f+\ell^2q^2)(t-t_0)}.
\label{eqn:Chi_derive}
\end{align}
A fit to the numerical data yields values for the 
potential $r$, 
the diffusion constant $D$, the 
bath temperature $T$,
and the intrinsic length scale $\ell$ as functions of time, see Fig. \ref{Fitting_params}. 
In these fits, we used the analytical expressions with initial time $t_0=0$ and with the initial correlations $C_q(0)$ extracted from the numerical FLEX data at $t=0$ in order to fit the numerical results for a larger time $t$.
As shown in Fig. \ref{Fitting_model_A} the functions can fit the form of the derivative quite well, while the fit of $C_{q}$ is dominated by the peak value and is not matching the tail of the function perfectly.
However, one finds that the obtained fit parameter depend strongly on time, and moreover, the parameters $\ell$ and $D$ which are usually assumed to be slowly varying at the phase transition vary strongly. As explained in the main text, this is expected, because the effective theory should not describe the early non-thermal electron phase of the evolution.
Getting the fit form for some linear ramp r(t) and T(t) would be feasible, but comes at the cost of additional parameters, making the analysis arbitrary.
However, trends like a rising temperature of the electronic background seem plausible as energy is injected into system. 
\begin{figure}
\includegraphics[scale=1]{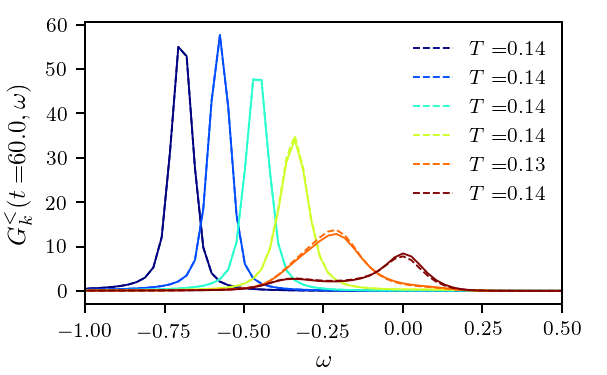}
\includegraphics[scale=1]{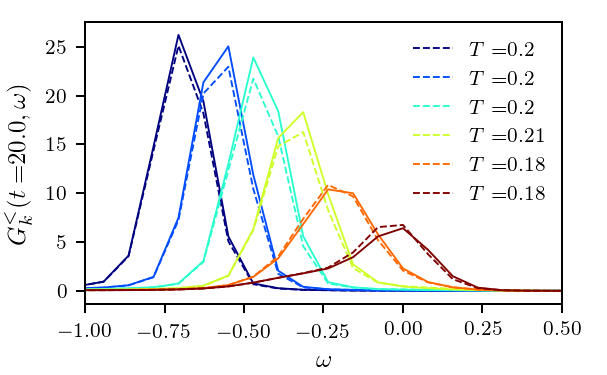}
\caption{$G^<_k(t,\omega)$ as function of $\omega$ for selected $k$ around the Fermi-edge (solid) and the corresponding value for the FDT based on the extracted temperature $T$ (dashed) for $t=60$ (\textit{upper panel}) and $t=20$ (\textit{lower panel}).}
\label{fig:Temp_verify_60}
\end{figure}

\subsection{Determination of the electron temperature} 

We determine the final temperature of the electronic system based on the detailed balance relation $\kappa_k(t,\omega)=\ln[G_k^<(t,\omega)/G_k^{>}(t,\omega)]$ from the Wigner-transformed Green's function. For a system in thermal equilibrium the function $\kappa(\omega)$ is given by:
\begin{equation}
\kappa(\omega)=-\omega/T+\Delta\mu,
\end{equation}
where the correction for the chemical potential $\Delta \mu=0.1$ is included to compensate the shift of the Fermi-edge due to the quenched interaction. The evaluation of this equation is only meaningful close to the 
Fermi-energy, 
where both $G^>_k$ and $G^<_k$ have values well above the numerical error. 
The extracted temperatures for the different electronic modes are then inserted in the Fermi-function $f(T,\omega)$ of the fluctuation dissipation theorem (FDT)
\begin{equation}
G^<_k(t,\omega)=A_k(t,\omega)f(T,\omega),
\end{equation}  
to verify that the system is thermalized. At $t=60$ the frequency resolution is sufficient to extract the spectra shown in the upper panel of Fig.~\ref{fig:Temp_verify_60}, where the individual mode temperatures lie within a narrow corridor around a global temperature $T\approx0.14$. For smaller times, see lower panel of Fig.~\ref{fig:Temp_verify_60}, there is no global mode temperature, indicating that the electrons are not fully thermalized at $t=20$. 
%\bibliographystyle{apsrev4-1}
%\bibliography{sources}

\end{document}